# Developing strategies to produce better scientific papers: a Recipe for non-native users of English

Osvaldo N. Oliveira Jr. [1,2], Valtencir Zucolotto[1] and Sandra M. Aluísio[2,3]

1. Instituto de Física de São Carlos, USP, CP 369, 13560-970 São Carlos, SP, Brazil
2. Núcleo Interinstitucional de Lingüística Computacional, NILC, CP 668, 13560-970 São Carlos, SP, Brazil
3. Instituto de Ciências Matemáticas e de Computação, USP, CP 668, 13560-970 São Carlos, SP, Brazil

e-mails: chu@ifsc.usp.br, zuco@ifsc.usp.br, sandra@icmc.usp.br

## Abstract

In this paper we introduce the AMADEUS strategy, which has been used to produce scientific writing tools for non-native users of English for 15 years, and emphasize a learn-by-doing approach through which students and novice writers can improve their scientific writing. More specifically, we provide a 9-step recipe for the students to compile writing material according to a procedure that has proven efficient in scientific writing courses.

## 1. Introduction and Background

The need to write papers in English has been a major problem for scientists around the world whose native tongue is not English. In order to exploit a strategy based on the reuse of written material, over the years we developed several computational writing tools with varied levels of sophistication in the use of artificial intelligence methods - the overall environment is called AMADEUS (AMiable Article DEvelopment for User Support) [1-9]. AMADEUS is aimed at catering for distinct users' needs, providing users with linguistic input for writing correct sentences and passages employed for specific purposes in a scientific paper, and is known as "appeal for assistance" [10] in language learning research. From the computational point of view, the reuse of linguistic material has been successfully employed in systems such as report generators, case-based letter generators, hypertext-based support systems for software documentation, and document drafting tools [11-17].

The Reference tool [1-4] was the first to be developed and works as a lexical database consultation where the units, instead of words, are expressions and standard sentences annotated according to the schematic structure of paper sections: abstract, introduction, review of the literature, methodology, results, discussion and conclusion. The access to these units is controlled by the user who uses them when describing tables and figures, comparing results and providing support for argumentation, for example. It was developed for domains in Experimental Physics and Computer Science. Later on its functionality was included in the Support Tool, described below, allowing access to a more detailed schematic structure of Introductions. The Reference Tool was aimed mainly at (1) researchers who are familiar with the academic writing but still need some feedback, (2) proficient or near-proficient English users who need to write under the constraint of time. Tests with this tool performed at the University of São Paulo, in São Carlos, showed that students' writing was improved significantly for they were able (1) to get started with their writing task overcoming their initial mental block, (2) to familiarise with the schematic structure of scientific articles, and (3) to produce a draft with good pieces of writing. However, the tool was not successful in helping less experienced writers, mainly because such users had difficulties in localising expressions, collocations and cohesive links appropriate to their needs.

For the latter type of user, we developed the Support Tool [5,6,18], which required a comprehensive corpus analysis because - despite its relatively well-defined schematic structure (see refs., [19-21]) – a scientific paper can be organised in different ways. Owing to the complexity of the linguistic analysis, this module covers only the Introductory sections of papers in Physics. The detailed schematic structure comprises 8 components, 30 rhetorical strategies and 45 different types of message (standard linguistic expressions). The tool was implemented as a case-based system (CBR) (see ref. [22]), with a case base of 54 annotated Introductions, with each case consisting of the Introduction plus its rhetorical structure. These knowledge sources were applied to the three stages of the writing process, i.e. planning, composing, and revising according to the writing model of Hayes-Flower [23]. Because it provides the rhetorical strategies for creating a paragraph, the Support Tool was effective in helping users to ensure cohesion and coherence of small chunks of text. The self-reviewing stage of the Support Tools is based on a Systemic Functional Linguistic (SFL) interpretation of textual revisions [24].

A limitation identified in the use of the Reference and Support tools was that students sometimes required feedback to their choices, which prompted us to develop the Critiquing Tool [7] and the CAPTEAP module for assessing performances (described below). The Critiquing tool provides structural knowledge at the supraparagraph level, indicating the most appropriate sequences of the schematic components in the various sections of a paper. It was aimed mainly at (1) researchers who have some experience with the academic writing, (2) novice English users who have problems in writing to a particular audience and purpose. For coping with the latter problem — to address a specific purpose — 51 short papers published in the CHI'96 Conference were classified according to the types of paper: empirical, experience, system, theory, and methodology papers. They were annotated with components from another structure (specific components structure) to allow a more fine-grained recovery suitable to the user needs. The tool works collaboratively in cycles — the user presents a product (the structure of a paper) to the system and the system gives feedback through critiques for improving the product. It employs two types of analysers to evaluate the product. The first evaluates it through guidelines while the other employs similarity metrics to compare cases recovered with the user's product. Five types of critique are generated: compliments, direct criticism, indirect criticism, direct suggestions and instructions. Furthermore, general suggestions for improving the user's structure are also provided.

AMADEUS suite also includes a module, CAPTEAP [25], which serves two purposes: it may function as a self-assessment tool, thus allowing users to monitor their progress, and as a summative tool for English proficiency tests (EPTs). The module was devised to assess skills novice researchers really need: (1) to understand and produce technical writing in English, (2) to recognise the schemata for the academic discourse. Accordingly, the tests involve questions addressing the role of each section in a paper, strategies for writing the more detailed structure of sections, summarisation operations, in addition to topics of the English language that are known to bring difficulties to non-native English authors. As far as the assessment of learning of English for Academic Purposes is concerned these tests are innovative because they use the Bloom's Taxonomy of Educational Objectives [26] to classify types and difficulty levels of proposed questions. It also uses the Admissible Probability Measures [27] as a score system to classify a student as informed, misinformed, partially informed, and uninformed in relation to a question item, instead of using the traditional right-wrong scoring system. The knowledge (scientific papers annotated with the schematic structure, writing strategies, for example) for the tests is taken from the Support and Critiquing tools' case bases.

As far as the writing processes are concerned, AMADEUS may conceptually be considered as consistent with the Flower and Hayes' model, for the activities while using the environment encompass several of the cognitive processes embedded in their model. In particular, AMADEUS is directly related to one crucial function predicted by Hayes' extended model [28], namely "reading to understand the task", which is achieved upon providing AMADEUS' users with linguistic input as well as cases of real text. The overall conception of the tools may be considered as belonging to the constructivist approach to learning - i.e. it emphasizes learning by doing.

In this relatively long period of time, we realized that language problems were often mixed up with difficulties in understanding the schemata and discourse of scientific papers, especially for less experienced scientists and students. In this context, these difficulties in producing good-quality texts for a paper are faced also for native English authors. This is a view shared by distinguished scientists that are not necessarily involved in scientific writing [29].

While acknowledging the vast literature in scientific writing and in acquiring English as a second language [8,9,30-32], we felt that students could benefit from straightforward guidelines with no need to learn the jargon of either fields. For such guidelines we focus on two main aspects, namely the paper viewed as a product of the research and recommendations for the students to build their own material to be used in the learning process, in a learning-by-doing approach.

## 2. Scientific paper as a product

In spite of the importance of getting papers published as the main indicator of the scientists` performance, students rarely get training in scientific writing. The basic features of a paper are learnt by intuition, which may be ineffective and/or inefficient. As a result misconceptions appear, leading to papers that are poorly structured and lacking clear focus. In Figure 1, we present a flowchart containing the main processes that we believe occur in conceiving, performing the research and getting the outcome published. Central in this chart is that ideas are what scientists should prioritize in their writing the paper. For novice scientists tend to structure their paper as if they were primarily publishing the results, rather than novel ideas and concepts that are ultimately supported by the results. The processes in Figure 1 include the design and implementation of the research based on the initial ideas. Once the results are obtained, we take the view that the authors should identify the main ideas on which the paper would be based. Note that these **ideas** may be different from the original ones, as often happens with the work proceeding along lines other than initially planned. Therefore, in case the final ideas differ from the original ones, the latter could only be mentioned as motivation or be dropped altogether. Indeed, if this distinction is not made, the paper will probably lack focus. The misconception associated with publishing results is mostly reflected in the Abstract, which sometimes brings a description of several results without emphasizing the main contributions of the paper.

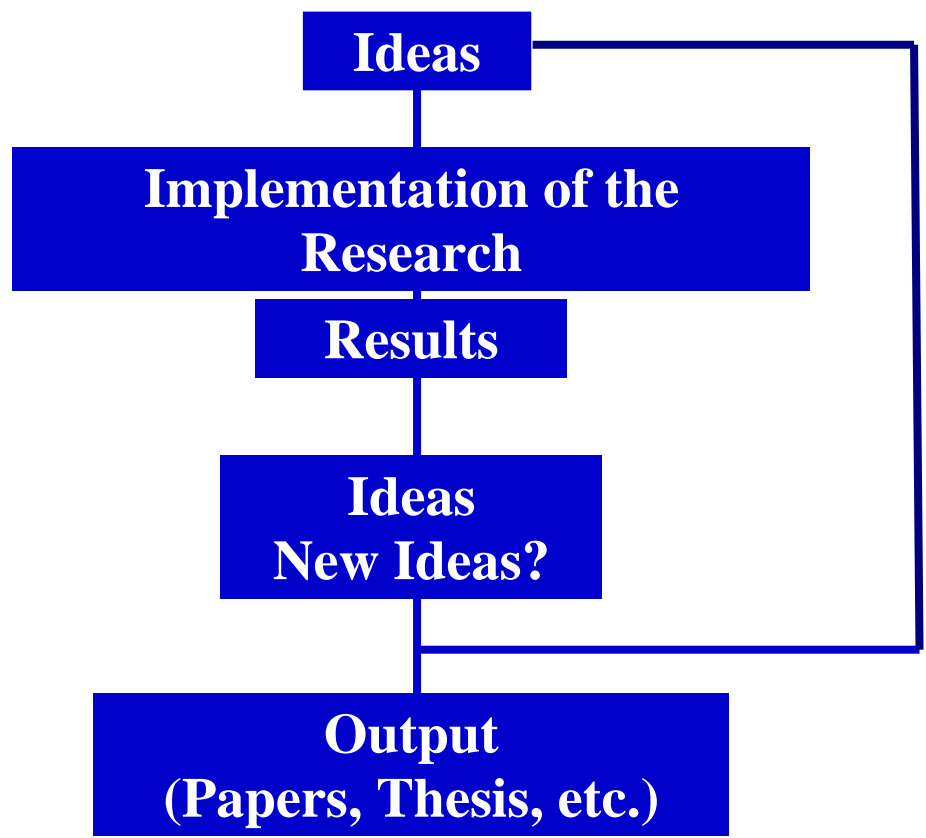

**Fig. 1 – Flowchart illustrating the main processes in designing and implementing a research project, also including the reporting of the main findings**

## 3. Guidelines for preparing reference material

Because we have evidence that compiling and using reference material is an efficient way of learning the scientific discourse, we present next a procedure for the student to develop his/her own material. We should remind that this strategy has nothing to do with **plagiarism – whole sentences are not reused; no factual information is reused.** The procedure consists in 9 steps, as follows:

**1st step:**

Select well-written texts from reliable sources and produced by native speakers. Read the material critically, annotating expressions that convey important messages and may be useful to reuse in the future.

**2nd step:**

Compile the expressions and sentences, clearly marking the reusable parts. The non-reusable parts are the gaps to be filled in. This procedure should be part of your learning life – never stop doing it.

**3rd step:**

Classify the materials according to the schemata of a scientific paper, such as that shown in ref. [5].

For this $3^{rd}$ step, two distinct strategies can be used: In the first, assign the expressions to the pre-defined scheme for the various parts of an article, together with the selection (e.g. an expression taken from a component from the Introduction is automatically classified as such). This strategy is advantageous because it is easy and quick, but has the disadvantage in that the user does not practice reshuffling the material. In the second strategy, select a large number of expressions (hundreds!) and only classify them later. It has the advantage of being more efficient to learn how to reuse the expressions, but it is more time-consuming.

**4th step:**

Practice filling in the gaps with your own material and/or based on other examples.

| | Example: | This …… . | | the | ……  |
|---|---|---|---|---|---|
| | | paper | addresses | | problem |
| | | letter | analyzes | | case |

**5th step:**

Start playing with the pieces, identifying different combinations that appear in the original texts and creating your own combinations (the bricks are the same but the houses will be different). In this process, try to enrich the possibilities by selecting other expressions (2nd step), and keep practicing filling in the gaps (4th step).

**6th step:**

Start all over again with the selected expressions, now classifying them according to rhetorical messages (e.g. describe, contrast, confirm, define, compare, introduce, etc). The idea is to have a collection of expressions to be retrieved as you wish to state specific contents. Furthermore, keep selecting further expressions and filling in the gaps.

**7th step:**

Start working with full text passages, rather than only with separate sentences. Repeat the procedures of combining pieces, as in Step 5. Now is the time to learn using connectives efficiently. Compile a list of expressions with *however, in contrast, indeed, on the other hand, furthermore, nevertheless, since, because*, etc.

**8th step:**

It is time to produce a full section of a paper. Select the subcomponents, and implement them by reusing material from your earlier practices. Fill in the gaps, for which help may be obtained by retrieving material from the practices. Check the use of connectives and the text coherence

**9th step:** Editing the text

Check the section for typos and other surface errors. Eliminate unnecessary words. Check the consistency of the subcomponents and their inter-relationship. Analyze the contents for completeness and accuracy.

**Final Remarks**

In offering a recipe for students and novice writers to improve their scientific writing in English, we are perfectly aware of the various oversimplifications adopted in terms of standardizing both writing processes and the scientific discourse. We were encouraged to put this recipe forward, nevertheless, because over the years we have seen students benefiting enormously from the AMADEUS strategy. We emphasize, however, that in all cases it was the practice and dedication of the user that made it a success. For this strategy cannot replace solid learning of English, and only works for users with reasonably good reception of English. Nevertheless, though the AMADEUS strategy was not conceived as a tool to learn

English, users may improve their proficiency by practicing with the language in context. For example, the material compiled should be excellent source for checking use of prepositions, phrasal verbs, connectives and even vocabulary of the field in focus. Finally, we remind our readers that using AMADEUS strategy should not be confounded with plagiarism, as we advocate the critical reading of written texts, from which non-factual material can be extracted as a way to learn standardized linguistic expressions that the users need to apply themselves.

## Acknowledgments

The authors are grateful to the many students and collaborators who helped develop AMADEUS over the years, and also acknowledge the financial support from CNPq (Brazil).